\documentclass{ametsocV6.1}
\usepackage{graphicx}
\usepackage{ascmac,amsmath,amssymb,mathtools, bm}
\usepackage{siunitx}
\usepackage{lineno}
\title{A Stochastic Lattice Model for Convective Self-aggregation \\Incorporating Longwave Radiative Effect}

\authors{Takuya Jinno,\aff{a}\correspondingauthor{Takuya Jinno, jinno@sus.u-toyama.ac.jp}
Hiroaki Miura,\aff{b} 
}

\affiliation{\aff{a}{Faculty of Sustainable Design, University of Toyama, Toyama, Japan} \\
\aff{b}{Graduate School of Science, The University of Tokyo, Tokyo, Japan}
}

\abstract{Self-aggregation of tropical convection is a universal feature observed in a diverse range of atmospheric environments. Several preceding models conceptualized the self-aggregation of convection as a phase transition driven by collisions between cold pool gust fronts. However, self-aggregation may also be influenced by various physical processes, such as surface fluxes, radiation, and moisture perturbations in the planetary boundary layer, and it remains unclear which process plays a dominant role. In this study, we develop a simple stochastic lattice model for the pattern formation of deep convection, inspired by the two-dimensional Ising model. Here, in addition to the process of cold pool collisions, which have an effect of triggering new convection, we incorporate the process of clear-sky radiative cooling that has an effect of suppressing deep convection as an interaction between clouds. Our results show that by amplifying the intensity of the clear-sky radiative cooling effect, the transition from a quasi-uniform to an inhomogeneous cloud field can be reproduced. The model also successfully explains the dependence of self-aggregation on several model parameters, such as the experimental domain size and the characteristic size of cold pools. 
Furthermore, by varying the distance over which the subsidence induced by radiative cooling extends, we succeed in capturing a pattern formation that closely resembles the convective clusters observed in the real atmosphere and three-dimensional numerical model simulations.}

\begin{document}
\nolinenumbers
This Work has been submitted to Journal of the Atmospheric Sciences. Copyright in this Work may be transferred without further notice.
\maketitle

%
%
%
%
%
%

%


\section{Introduction}
The cloud self-aggregation observed in radiative-convective equilibrium (RCE) experiments using cloud-resolving models (CRMs) is initiated and maintained through the complex interplay of various physical processes, including surface fluxes, radiation, moisture perturbations in planetary boundary layers, and convective entrainment and detrainment \cite[e.g.][]{Coppin_2015, Windmiller_2019}. A defining characteristic of cloud self-aggregation is the emergence of macroscopic structures at specific spatial and temporal scales through the collective behavior of microscopic clouds. To elucidate the relationships among multiple physical processes and the mechanisms of self-aggregation as a form of pattern formation, it is essential to account for interactions between clouds. These cloud-cloud interactions have been extensively studied by incorporating them into a variety of theoretical and numerical models.

One way to investigate the interactions between clouds is to represent large-scale cloud organization as a bifurcation of steady solutions of nonlinear equations. In this approach, the behavior of self-aggregation has been analyzed as a dynamical system. \citet{Haerter_2019} developed a theoretical model that adopted an assumption that the effect of precipitation-induced cold pools scaled quadratically with local number density in the vicinity of existing convection cells. It was shown that a continuous phase transition from unaggregated to aggregated regimes occurred depending on the strength of interaction among clouds through the cold pool collisions. \cite{Khouider_2019} investigated the chaotic dynamics of the cloud work function, the cloud base mass flux, and the cloud area fraction on the horizontal scale of \SI{100}{km} by relaxing the quasi-equilibrium assumption of \citet{Arakawa_1974} to allow the system to have memory of a finite time scale. As a result, multiple equilibria of cloud area fraction having various regimes of the cloud organization were demonstrated, depending on the strength of the interaction between different cloud types. 

The other way is to construct numerical models that represent spatial distribution of clouds or moisture by taking into account the amplitude of interactions based on the distance between simulated clouds.
The self-aggregation of clouds is characterized by a horizontally heterogeneous distribution of thermal energy and water vapor, which are separated into dry and moist regions. Not only in atmospheric convection but also in a wide variety of emergent systems, it is known that the necessary elements for such spatial separation are the existence of bistable equilibria and local interactions \citep{Muller_2022}. With these two elements, the spatial distribution of some physical variables in the system can be separated into two stable phases, which correspond to the moist and dry equilibria in the tropical atmosphere \citep{Sobel_2007,Sessions_2010}. It is observed that the characteristic size of each moist and dry cluster grows with time, and this tendency is called coarsening \citep[p. 49]{Hutt_2020}.

One of the simplest models that represent coarsening is the reaction-diffusion equation
\begin{equation}
  \label{TDGL}
  \frac{\partial q(\mathbf{r})}{\partial t} = -\frac{\delta F_L}{\delta q} = -\frac{\delta V}{\delta q}+D\nabla^2 q(\mathbf{r})
\end{equation}
where $q$ is humidity content, $F_L$ is Landau free energy, $V$ is some kind of potential function of $q$, and $D$ is a diffusion constant. Eq. (\ref{TDGL}), which is called the time-dependent Ginzburg-Landau equation, describes the evolution of $q$ at a location $\mathbf{r}$ as a derivative of $F_L$. By setting $V$ to have a double-well structure that depends on $q$, the humidity content at each location has bistable equilibrium and the diffusion process represented by the second term on the right-hand side induces self-aggregation with coarsening as a mesoscale structure. \citet{Craig_2013}, \citet{Bretherton_2005}, and \citet{Windmiller_2019} developed the reaction-diffusion model and reproduced cloud self-aggregation as coarsening of the moisture field. These previous studies formulated the local interaction among clouds in different ways, but they commonly employed an assumption to establish a double-well structure for the form of the potential $V$.

In addition to the reaction-diffusion models, two-dimensional grid models that explicitly simulate the expansion and collision of cold pool gust fronts have also been proposed.
For example, the models by \citet{Haerter_etal19} and \citet{Boing_2016} represented cold pool collisions and the development of new convective cells as a result. They also showed that the effect of cold pool dynamics acts to promote the aggregation of convective cells.

Past studies on the development of the simplified numerical models have focused primarily on the local effects via cold pool outflows developed around convective clouds, as described above. In the existing object-based numerical models for the cold pool interaction such as \citet{Haerter_2019} and \citet{Boing_2016}, the remote effect of longwave cooling in the subsidence region is not incorporated as an interaction between clouds. However, it is known that there are also interactions that work remotely between clouds through the large-scale field \citep{Wing_2014,Emanuel_2014}. In the dry regions, subsidence needs to be strengthened to generate adiabatic heating in the upper troposphere that compensates increased longwave radiative cooling in the lower troposphere. This process increases the convective stability of the atmosphere and suppresses the generation of new convection cells in the cloud-free regions. This clear-sky subsidence couples with the lower-tropospheric convergence of moisture in the moist regions. The connection between radiation and large-scale circulation is suggested to explain the characteristic length of aggregated cloud clusters \citep[e.g.][]{Coppin_2015, Yanase_2020}. Therefore, formulating the effect of subsidence as a feedback from the smaller-scale convection to the larger-scale field will lead to a better understanding of the interactions between scales in the atmospheric system.

In this paper, we propose a framework of the lattice model that enables a unified representation of the interactions of convection across different scales. In section \ref{model} the details of the stochastic lattice model is described. Our model is based on the one presented by \citet{Randall_1980}. Their original model was designed to reproduce cloud fields that exhibit various characteristics such as squall lines, non-squall bands, open cells and cloud streets, which are on spatial scales of 10-100 km and time scales of 2-10 h. This study formulates the effects of precipitation cold pools and subsidence drying so that the model can reproduce the self-aggregation of deep convection observed in RCE experiments. This model configuration enables one to investigate how the relative intensity of multiple interaction processes between clouds determines the transition from the phase of random convection to the phase of aggregated convection. In section \ref{result}, we present results on changes in the quasi-steady state in response to model parameters, along with a theoretical interpretation of the model's behavior from the perspective of bifurcations in steady-state solutions. It is shown that the amplitude of the subsidence effect, the characteristic size of cold pools, and the size of the experimental domain play important roles in determining whether the separation of moist and dry regions occurs. 
Additionally, by varying the distance over which the subsidence induced by radiative cooling extends, we obtain indications regarding the spatial pattern formation of cloud clusters.
Discussion and conclusions are given in section \ref{conclusion}.

\section{Model description}
\label{model}
This section presents a stochastic model that represents local and remote interactions between convective cells. The underlying model was developed by \citet{Randall_1980} and we modified it to better represent the physics of the self-aggregation of deep convection.
In this study, the focus is on conceptualizing the effects of the spatial distribution of an ensemble of convection, which may be determined by relative positions between all of the pairs of clouds. To ensure that the model is simple, the statistical variations of cloud properties such as the intensity or the size are neglected.

The model describes the time evolution of two variables on a horizontal two-dimensional grid space. One is the two-state variable $I(\mathbf{r})$ that indicates whether a convection occurs or not at a position $\mathbf{r}$. 
The other is $J(\mathbf{r})$ that represents the amplitude of boundary layer turbulence that triggers convection, which is offset so that the threshold of convection initiation becomes zero. It is interpreted as the degree to which a convection actively develops at a position $\mathbf{r}$.
We assume that deep convection is generated among convectively active grid points at a prescribed probability $p_{\gamma}$ , which is formulated as
\begin{equation}
  \label{eq_I}
    I(\mathbf{r}) =
    \begin{cases}
      1 & \mathrm{if\ \ } J(\mathbf{r}) > 0 \mathrm{\ \ and\ \ } \gamma(\mathbf{r}) < p_{\gamma} \\
      0 & \mathrm{otherwise}.
    \end{cases}
\end{equation}
The variable $\gamma(\mathbf{r})$ is a stochastic variable assigned to each grid point with the uniform distribution over the range from 0 to 1.

The tendency of $J$ at a position $\mathbf{r}$ is assumed to be the sum of (i) the interaction between clouds, denote by $F(\mathbf{r})$, which is determined by the relative positions of convective cores around the position $\mathbf{r}$, (ii) the large-scale forcing due to non-cloud processes $\alpha$, which is assumed to be horizontally uniform regardless of spatial distribution of convection, and (iii) the linear dissipation of $J$ with a constant time scale $\tau_d$. Then, the prognostic equation of $J$ is described as
\begin{equation}
  \label{eq_prog}
    \frac{dJ(\mathbf{r})}{dt} =  F(\mathbf{r}) + \alpha - \frac{1}{\tau_d} J(\mathbf{r})
\end{equation}
for each grid point. 

\begin{figure}
 \begin{center}
  \includegraphics[width=120mm]{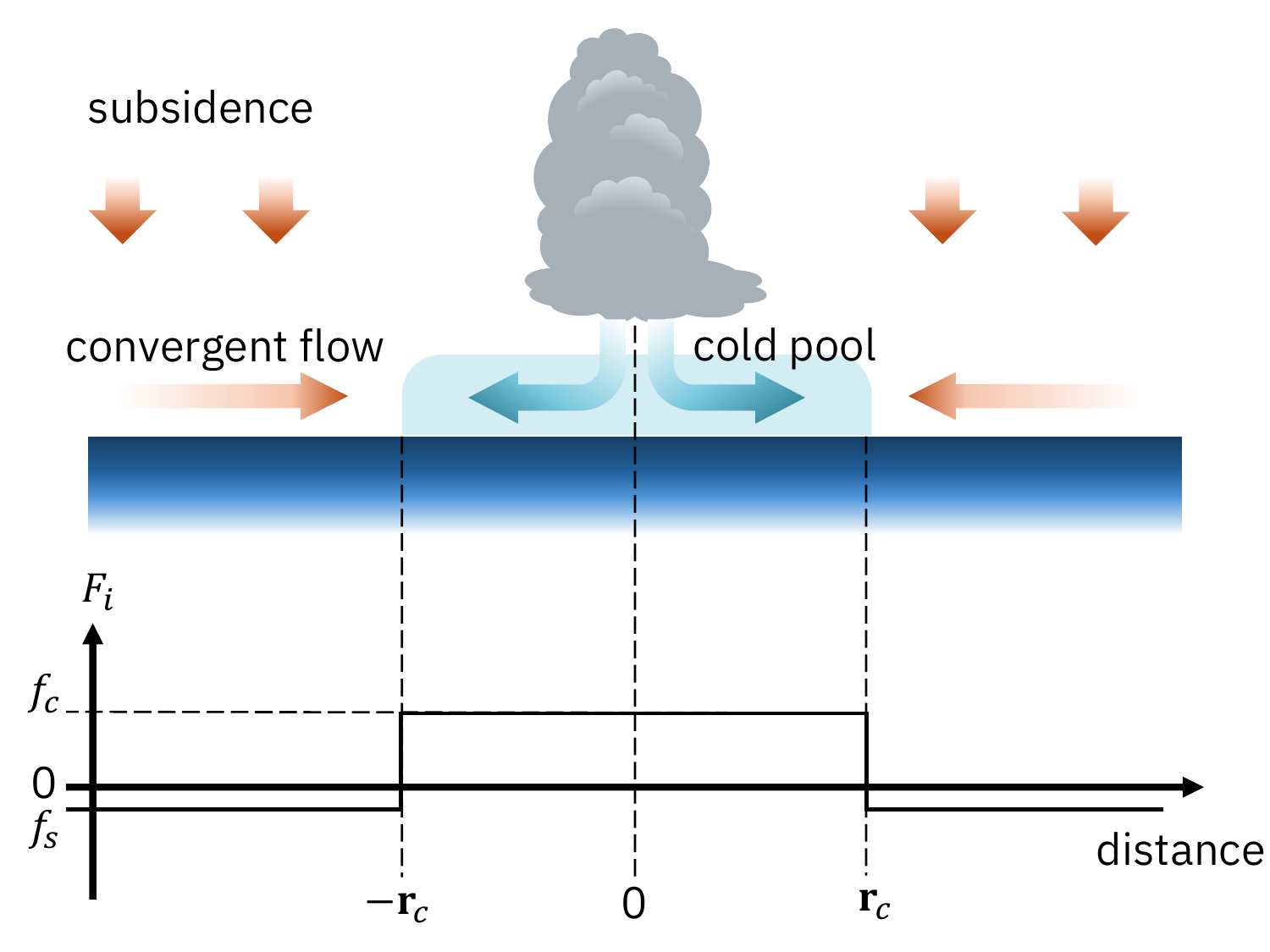}
  \caption{Schematic illustration for the promotion and suppression effects of a cloud on the surrounding field as a function of the distance from a convective site.}
  \label{stab-schem}
 \end{center}
\end{figure}

The interaction term $F(\mathbf{r})$ is given as follows. 
We assign a positive integer index $i$ to one of the grid points satisfying the condition $I(\mathbf{r})=1$ in Eq. (\ref{eq_I}) and refer to it as a convective grid point $\mathbf{r}_i$.
A convective grid point $\mathbf{r}_i$ superimposes the interaction effects illustrated in Fig. \ref{stab-schem} onto the surrounding field. We assume that $J(\mathbf{r})$ is increased if $\mathbf{r}$ is located within the radius $r_c$ from the convective grid point $\mathbf{r}_i$ since the spreading fronts of precipitation-driven cold pools have an effect to promote the generation of new convective cells. On the other hand, we assume that $J(\mathbf{r})$ is decreased due to the drying of subsidence if $\mathbf{r}$ is located farther than $r_c$ from the convective grid point $\mathbf{r}_i$. Thus, the magnitude of interaction effect from the $i$-th convective grid point $\mathbf{r}_i$ onto $\mathbf{r}$ is formulated as
\begin{equation}
  \label{eq_f}
    F_i(\mathbf{r},\mathbf{r}_i) =
    \begin{cases}
      f_c > 0 & (|\mathbf{r}-\mathbf{r}_i|< r_c) \\
      f_s < 0 & (|\mathbf{r}-\mathbf{r}_i|\geq r_c).
    \end{cases}
\end{equation}
At each time step, the interaction term is calculated at all model grid points as the sum of the influences from the convective grid points, $F(\mathbf{r})=\sum^N_{i=1}F_i(\mathbf{r},\mathbf{r}_i)$. Here, $N$ is the total number of convective cells in the model domain at each time step, which is equivalent to the number of grid points satisfying the condition of convection, $I(\mathbf{r}_i)=1$.

The grid resolution $dx$, which corresponds to the size of a single convection core is set to \SI{2}{km} for both directions of the two orthogonal axes. A square domain with doubly periodic lateral boundary conditions spans \SI{256}{km} by \SI{256}{km} in the horizontal dimensions. The domain edge length is denoted by $L$ in the following. The interval of the time levels is set to 1 hour, which is roughly on the same order of lifetime of a deep convective cell. The initial conditions of both $I$ and $J$ are zero everywhere. Each simulation is performed for \SI{500}{} time steps, which is sufficiently long to reach quasi-equilibrium states (justified later in Fig. \ref{JN-time}). Here, the default value of the parameter $r_c$ is set to \SI{20}{km}. The ``vicinity'' of convection $r_c$ is estimated to be \SI{20}{km} as a product of the time interval and the typical initial velocity of cold pool gust fronts $\sim 5\ \mathrm{m\ s^{-1}}$ \citep{Zuidema_2017, Romps_2016}. In the real atmosphere, cold pools reach maximal radii of \SI{10}{} to \SI{100}{km} over the ocean surface \citep{Haerter_2019}. Other parameters are summarized in Table \ref{tab_param}. They are determined so that the value of $J$ in quasi-steady state is on the order of unity.

We will conduct a series of sensitivity tests that varies the value of $f_s$ from $-0.0005$ to $-0.0045$ to investigate the convection suppression effect of the clear-sky subsidence on the self-aggregation. The values of $f_s$ are at least one order of magnitude smaller than the value of $f_c$.
Subsequently, to investigate the influence of other model parameters on self-aggregation, the results of sensitivity experiments in which the values of $r_c$ and $L$ are varied independently are also presented in Section \ref{result}. In Eq. (\ref{eq_f}), the effective range of the subsidence effect $f_s$ is set as $|\mathbf{r}-\mathbf{r}_i|>r_c$. Exceptionally, for experiments with larger domain sizes than the default ($L=\ $\SI{512}{km} and \SI{1024}{km}), this range is modified to $r_c<|\mathbf{r}-\mathbf{r}_i|<\ $\SI{200}{km}. This condition is introduced with the intention of examining how the spatial pattern formation of cloud clusters changes in response to the typical size of the subsidence region induced by a single convection cell.

\begin{table}[]
 \begin{center}
  \begin{tabular}{clc} \hline
    Parameter & Description & Value \\ \hline
    $f_c$ & Forcing in the vicinity of convection sites & 0.1\\ 
    $f_s$ & Forcing in the subsidence region & $-0.0005$ to $-0.0045$\\
    $\alpha$ & Spatially uniform effect of non-cloud processes & 0.1\\
    $\tau_d$ & Constant dissipation timescale [hour]& 2\\
    $p_{\gamma}$ & Seeding rate of convection & 0.01\\
    $r_c$ & Radius of local interaction from existing clouds [km] & 20 \\
  \end{tabular}
  \end{center}
  \caption{List of the model parameters and their values.}
  \label{tab_param}
\end{table}

Compared to the numerical model by \citet{Randall_1980}, the current model is modified to have a dynamical memory on a finite time scale $\tau_d$ by introducing the dissipation term in Eq. \ref{eq_prog}. This formulation allows us to analyze the behavior of the system with stochastic fluctuations around steady states, which correspond to the radiative-convective quasi-equilibrium in the cloud-resolving model simulations. This model also generates an intermittent and chaotic oscillatory behavior depending on the settings of certain parameters.
Three-dimensional RCE simulations have shown that a competition between the two lower-tropospheric divergent flows, which are respectively driven by cold pools and dry subsidence, determines whether dry regions can exist as a steady state \citep{Yanase_2020, Coppin_2015}. In our lattice model, the formulation of the interaction term (Eq. (\ref{eq_f})), which is simpler than that of \citet{Randall_1980}, proposes an advantage to allow for a linear stability analysis of the dry and moist solutions depending on the relative strength of the two processes. Results of a linear stability analysis and its relevance to the bifurcating behavior of our lattice model will be discussed in subsection \ref{result}\ref{stability}.

\section{Results}
\label{result}
\subsection{Model behavior of self-aggregation}
\label{var_fs}
In all simulations, the spatial distributions of $J$ and $I$ reach quasi-equilibria after about 100 steps and show different spatial patterns depending on the values of $f_s$. Stronger suppression of convection in the subsidence region, i.e., smaller values of $f_s$, result in the heterogeneous distributions of $I$ and $J$, which is similar to the self-aggregation observed in the RCE simulations.

\begin{figure}
  \begin{center}
    \includegraphics[height=190mm]{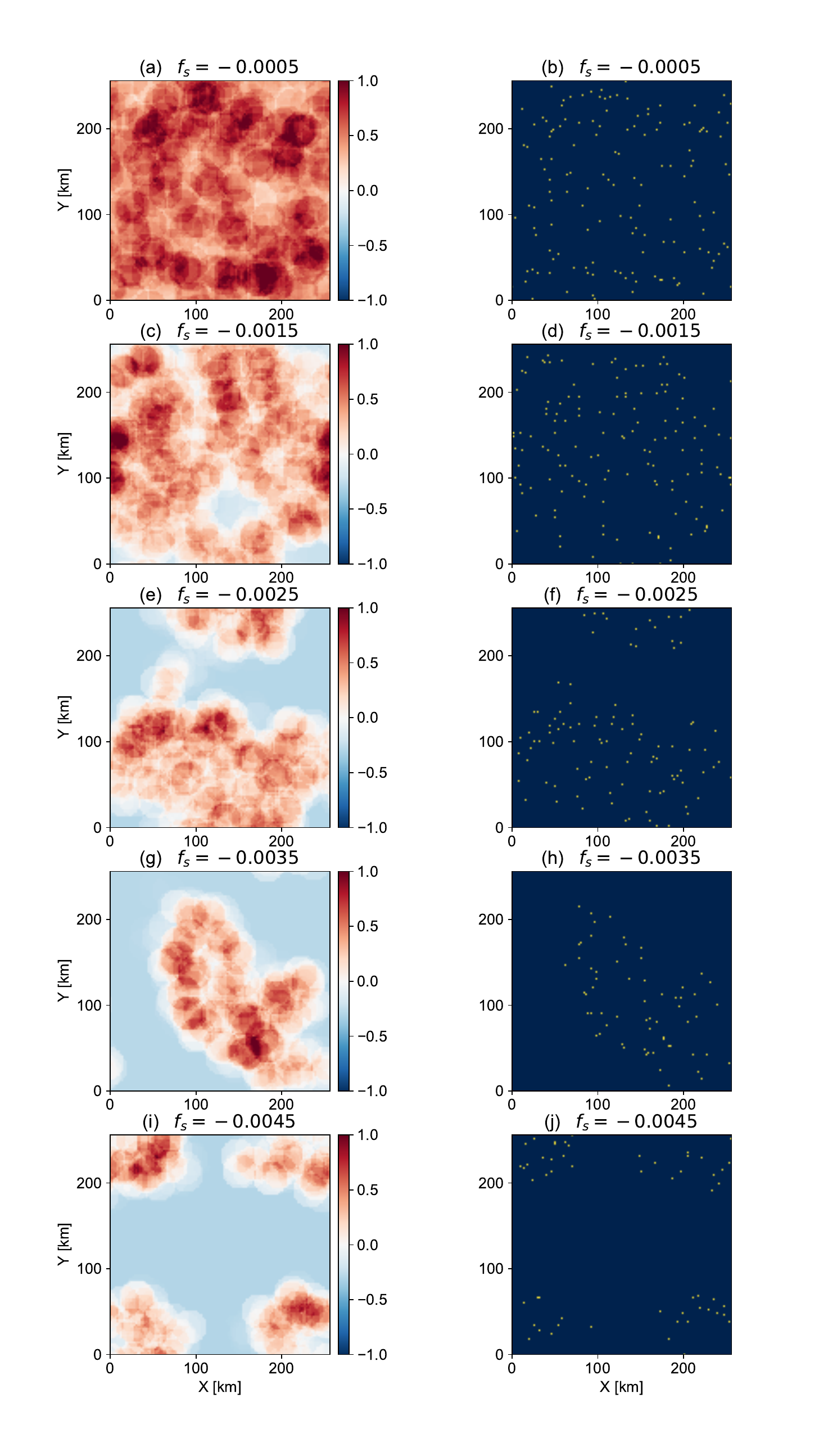}
    \caption{Snapshots of the horizontal distribution of $J$ (left column) and $I$ (right column) at $t=300$. The convective grid points ($I=1$) are marked as yellow dots in the $I$ fields. The values of $f_s$ are (a, b) $-0.0005$, (c, d) $-0.0015$, (e, f) $-0.0025$, (g, h) $-0.0035$, and (i, j) $-0.0045$, }
    \label{fr-imshow}
  \end{center}
\end{figure}

Figure \ref{fr-imshow} shows snapshots of $J$ and $I$ in each simulation after 300 steps. With $f_s=-0.0005$, $J$ is positive almost everywhere in the domain (Fig. \ref{fr-imshow}a), and convective events are distributed randomly at least apparently (Fig. \ref{fr-imshow}b). We can see a small portion of convectively stable region in $f_s=-0.0015$ (Fig. \ref{fr-imshow}c), but those small dry patches have a very short lifetime and disappears within several hours. In the three cases with smaller values of $f_s$ ($-0.0025$, $-0.0035$, and $-0.0045$), the cluster of convection is surrounded by a convection-free region (Fig. \ref{fr-imshow}e-\ref{fr-imshow}j), which persists stably longer than 10 days, with the edges of the clusters moving back and forth slowly.

\begin{figure}
  \begin{center}
    \includegraphics[width=0.75\linewidth]{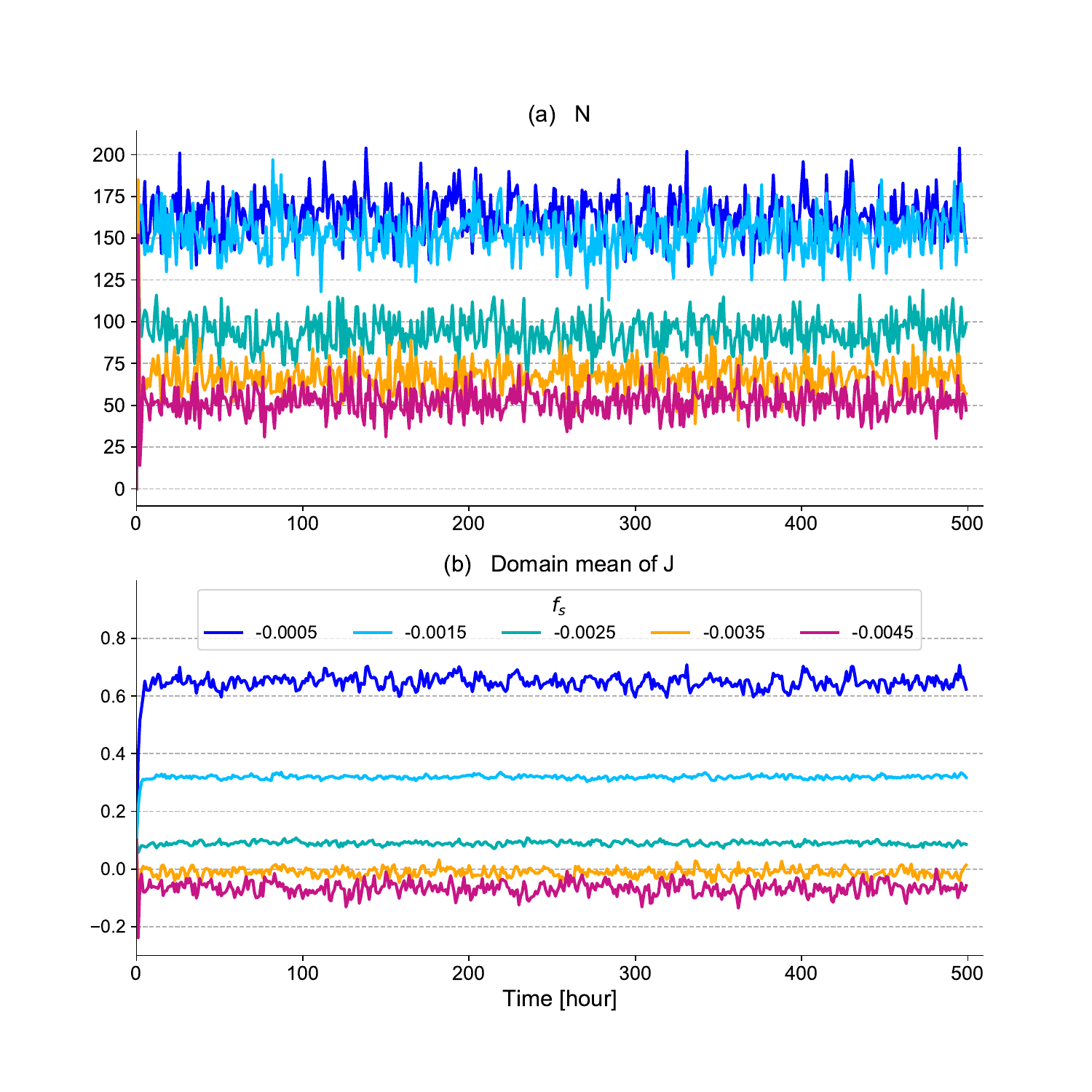}
    \caption{Time evolution of (a) the number of convective grid points and (b) the domain mean of $J$.}
    \label{JN-time}
  \end{center}
\end{figure}

The smaller $f_s$, the larger the area fraction of the subsidence region. In the subsidence region, the value of $J$ is negative, and it results in a smaller number of convective sites in the domain. The time evolution of domain averaged $J$ and $N$ are displayed in Fig. \ref{JN-time}. In the two cases of $f_s = -0.0005$ and $-0.0015$, the variations of $N$ is steadily in the range of about 150 to 170 (Fig. \ref{JN-time}a). This number is close to the expected value of $N$ ($p_{\gamma}L^2/dx^2=0.01\cdot128^2=163.84$) which is derived by assuming positive $J$ in the whole domain. This case corresponds to the regime where convective sites scattered over the entire domain. In the two aggregated cases of $f_s=-0.0035$ and $-0.0045$, the area average of $J$ in the quasi-steady state falls below 0, because of the increase in the area fraction of the subsidence region ($J<0$).
The region of $J<0$ is maintained because that location is too far from convective cells to receive the influence of cold pools.
In the simulation with $f_s=-0.0025$, there is a small but persistent subsidence region. We can regard  the quasi-steady state as an intermediate one between the aggregated and the scattered convection states.

\begin{figure}
  \begin{center}
    \includegraphics[width=0.65\linewidth]{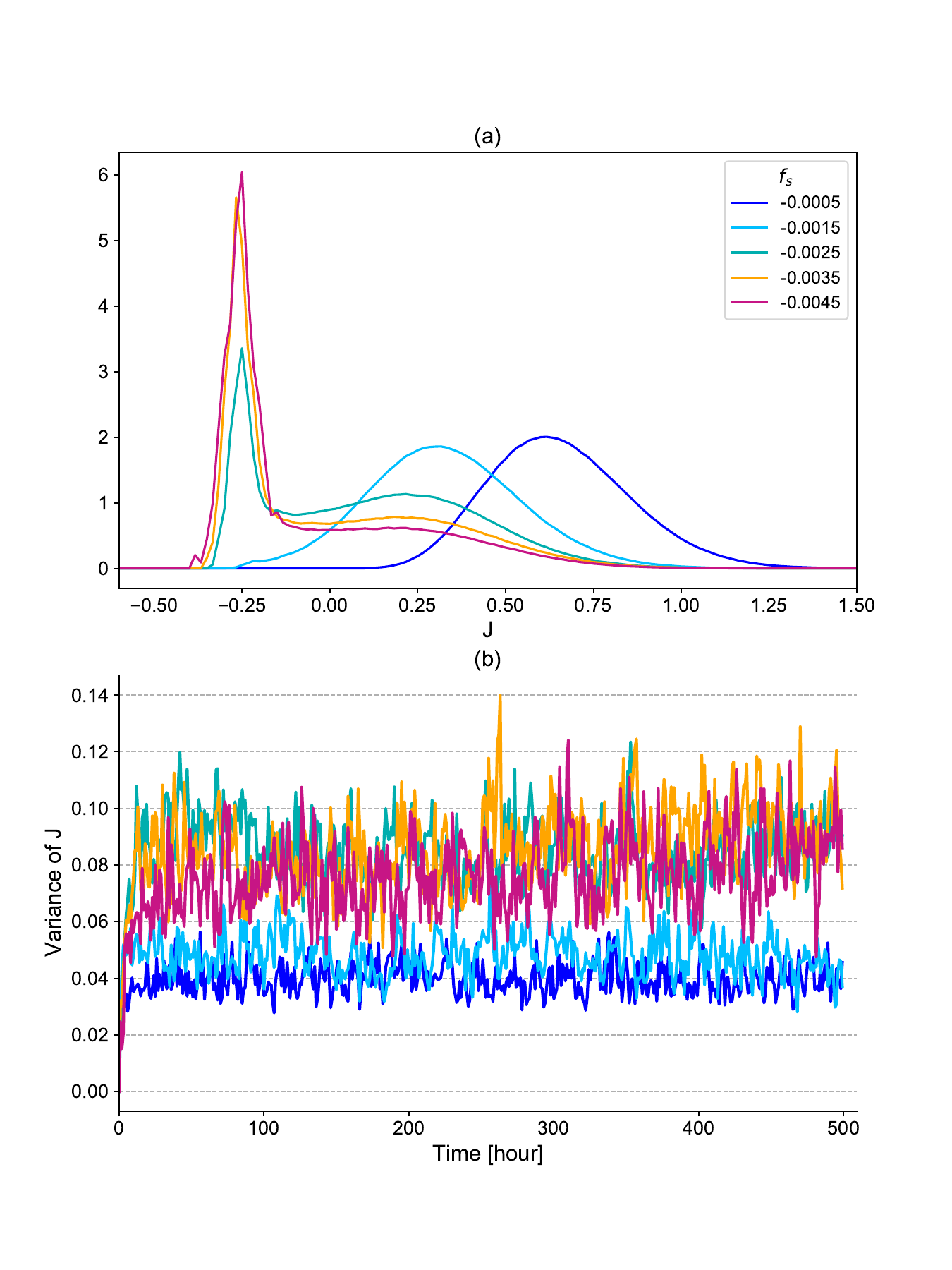}
    \caption{(a) PDFs of $J$ under the quasi-steady states. (b) Time evolutions of the variance of $J$.}
    \label{J-pdf}
  \end{center}
\end{figure}
\begin{figure}
  \begin{center}
    \includegraphics[width=0.7\linewidth]{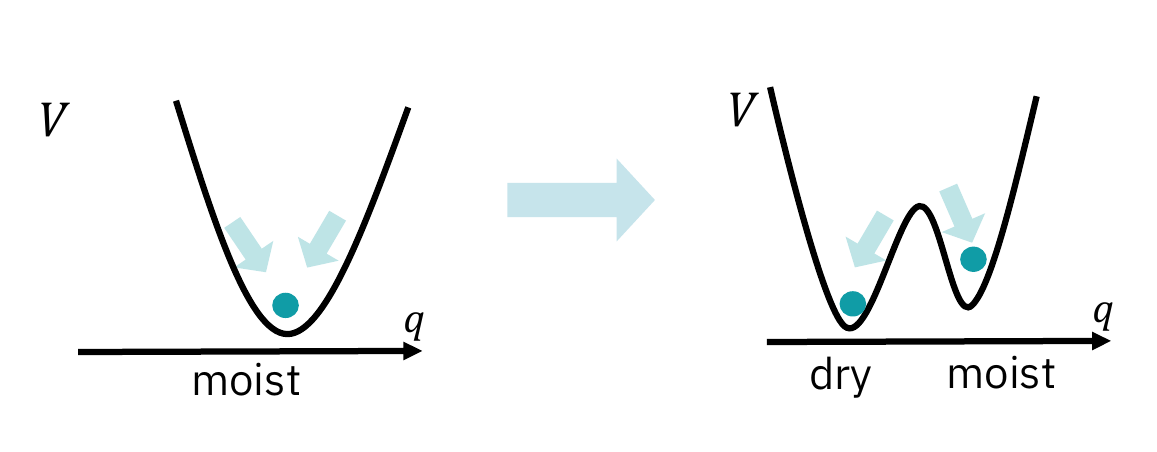}
    \caption{Schematics of the single-well (left) and the double-well (right) potentials. The double-well potential represents the bistability of moist content $q$. In the aggregated convection system (right), the value of $q$ evolves toward one of the potential minima depending on its initial value relative to the local maximum.}
    \label{bistable-schem}
  \end{center}
\end{figure}

In the cases where the self-aggregation solution is reproduced, the frequency distribution of $J$ becomes to have double peaks.
This is similar to the results of the RCE studies where the horizontal distributions of precipitable  water or convective available potential energy become double-peaked. Figure \ref{J-pdf} shows (a) the PDF of $J$ during quasi-steady state for  $t>100$ hours and (b) the time evolution of the variance of $J$. The two simulations of scattered convection, $f_s=-0.0005$ and $-0.0015$, have single-peaked PDFs. In the other three cases, there are two prominent peaks, one is negative and the other is positive. The variance of $J$ in the aggregated cases is larger than the scattered cases.
Considering the correspondence with the reaction-diffusion model Eq. (\ref{TDGL}), this double-peaked distribution function can be considered as the counterpart of the double-well shape of the reaction potential of water vapor $V(q)$ (Fig. \ref{bistable-schem}). A single stable solution corresponds to a state of unorganized convection ensembles while two stable solutions correspond to a regime with a dry convection-free state and a moist convectively active state.

\subsection{Dependence of the self-aggregation on the size of cold pools}
\label{var_a}
In this subsection, we investigate how the criteria for self-aggregation change depending on the parameter for the typical size of the cold pool ($r_c$). The radius of cold pools is suggested to be a key factor that determines the geometry and scale of self-aggregation events \citep{Haerter_etal19}. Here, the value of $r_c$ in Eq. (\ref{eq_f}) is set to \SI{10}{km}, \SI{20}{km}, \SI{30}{km}, and \SI{40}{km}, while $f_s$ is fixed at $-0.0045$ (the value of the aggregated case in the previous subsection) and the other parameters remain the same as in Table \ref{tab_param}.

\begin{figure}
  \begin{center}
    \includegraphics[width=0.9\linewidth]{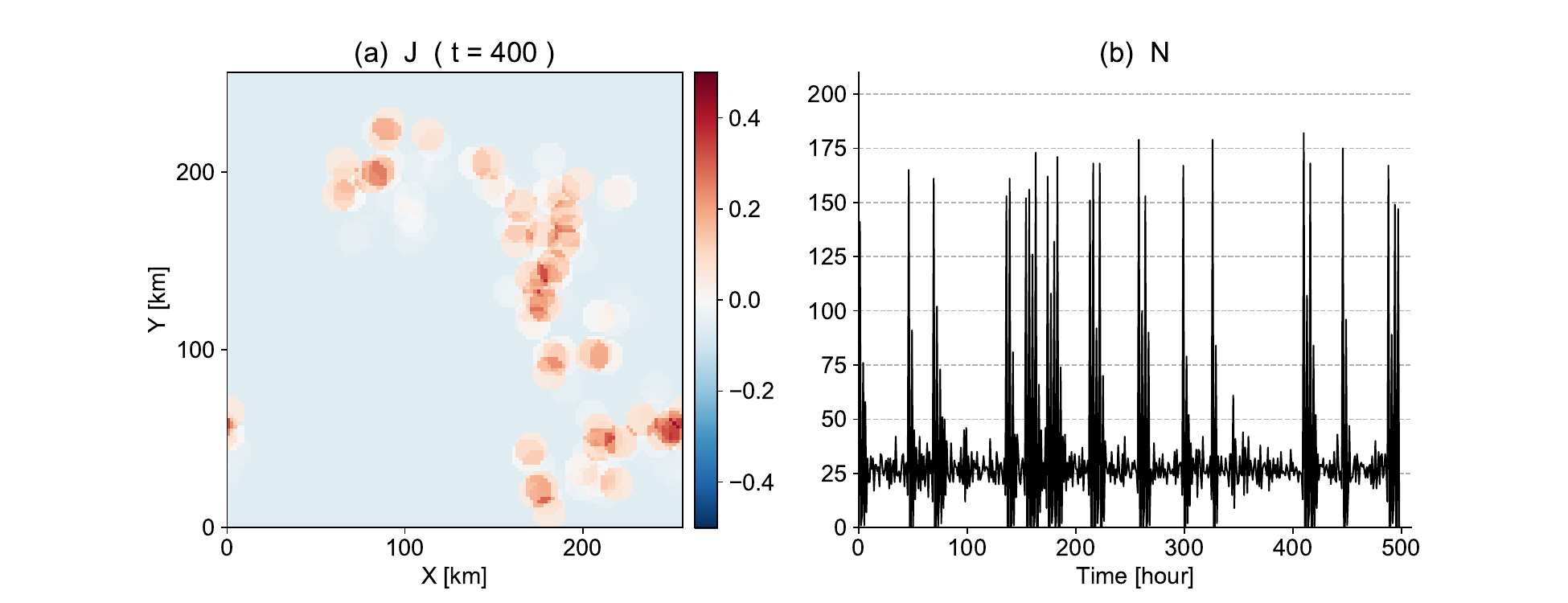}
    \caption{(a) Snapshot of the spatial distribution of $J$ at $t=400$ and (b) time series of number of convective grid points $N$ in the simulation with $r_c =\SI{10}{km}$.}
    \label{rc10km}

    \includegraphics[width=0.9\linewidth]{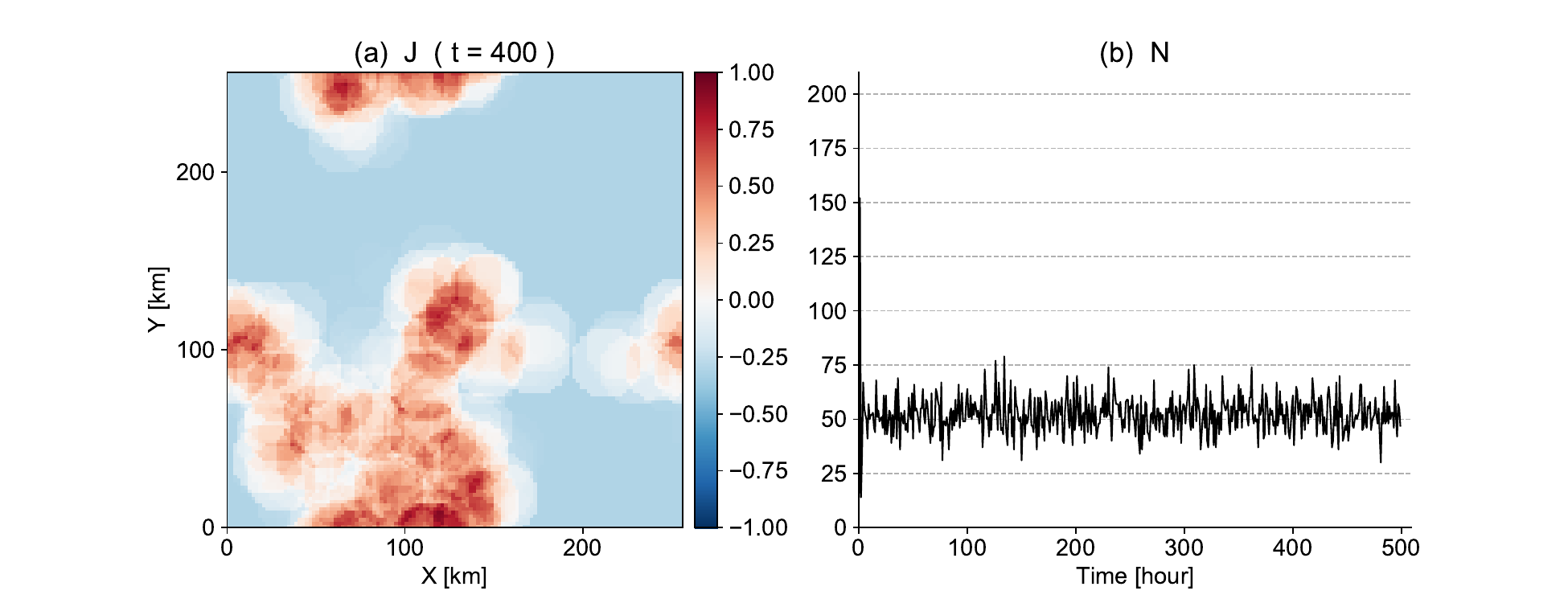}
    \caption{Same as Fig. \ref{rc10km}, but for the simulation with $r_c =\SI{20}{km}$.}
    \label{rc20km}
  \end{center}
\end{figure}
\begin{figure}
  \begin{center}
    \includegraphics[width=0.9\linewidth]{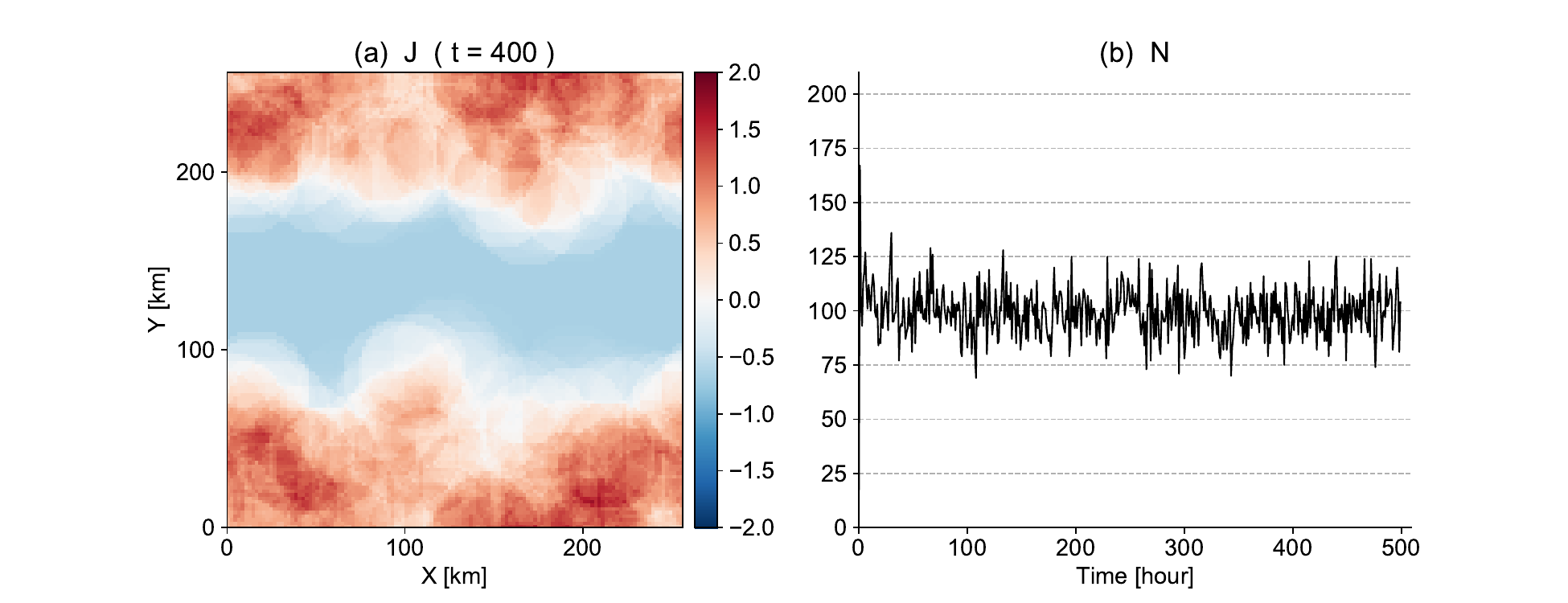}
    \caption{Same as Fig. \ref{rc10km}, but for the simulation with $r_c =\SI{30}{km}$.}
    \label{rc30km}

    \includegraphics[width=0.9\linewidth]{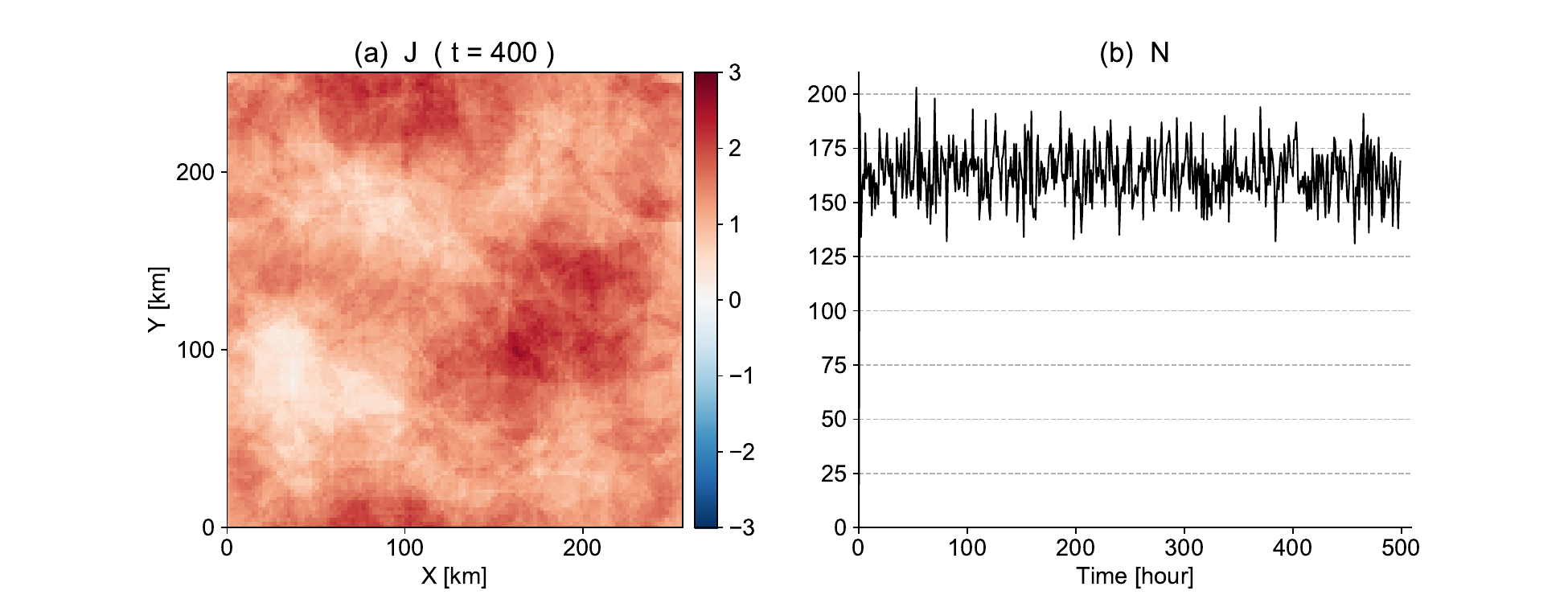}
    \caption{Same as Fig. \ref{rc10km}, but for the simulation with $r_c =\SI{40}{km}$.}
    \label{rc40km}
  \end{center}
\end{figure}

The time series of the number of convection and the domain average of $J$, and the horizontal distribution of $J$ in quasi-steady state shows noticeably different features depending on the size of the cold pool (Figs. \ref{rc10km} to \ref{rc40km}).
In the experiment with the cold pool radius of $r_c =\ $\SI{10}{km}, a relatively small number of convective cells are generated in close proximity (Fig. \ref{rc10km}a). The time series of the number of convection and domain mean of $J$ show a chaotic behavior with intermittent oscillatory events that generate numerous convection simultaneously (Fig. \ref{rc10km}b). As the radius of the cold pool is increased, the organized convective clusters become larger (Figs. \ref{rc20km}a, \ref{rc30km}a, and \ref{rc40km}a), and the number of convective grid points increases (Figs. \ref{rc20km}b, \ref{rc30km}b, and \ref{rc40km}b).
In the experiment with a cold pool radius of \SI{40}{km}, the whole domain is covered by the convectively active region ($J>0$) and the persistent subsidence branch in the large-scale circulation is not maintained (Fig. \ref{rc40km}). This suggests that the size of the cold pool is one of the factors that determines whether the potential $V(q)$ has a local minimum for the dry state (right panel of Fig. \ref{bistable-schem}) and consequently the self-aggregation is reproduced.

\subsection{Dependence of spatial patterns on the domain size}
\label{var_A}
Here, the relationship between the model domain size and the spatial pattern of cloud field is examined. It is known that for RCE simulations, the geometry of the simulation domain is supposed to be one of the major factors to determine whether the convective self-aggregation occurs or not \citep{Muller_2012, Yanase_2020}. Here, we examine how the quasi-steady states of the current model depend on the different domain sizes when the value of $f_s$ is fixed.

\begin{figure}
  \begin{center}
    \includegraphics[width=130mm]{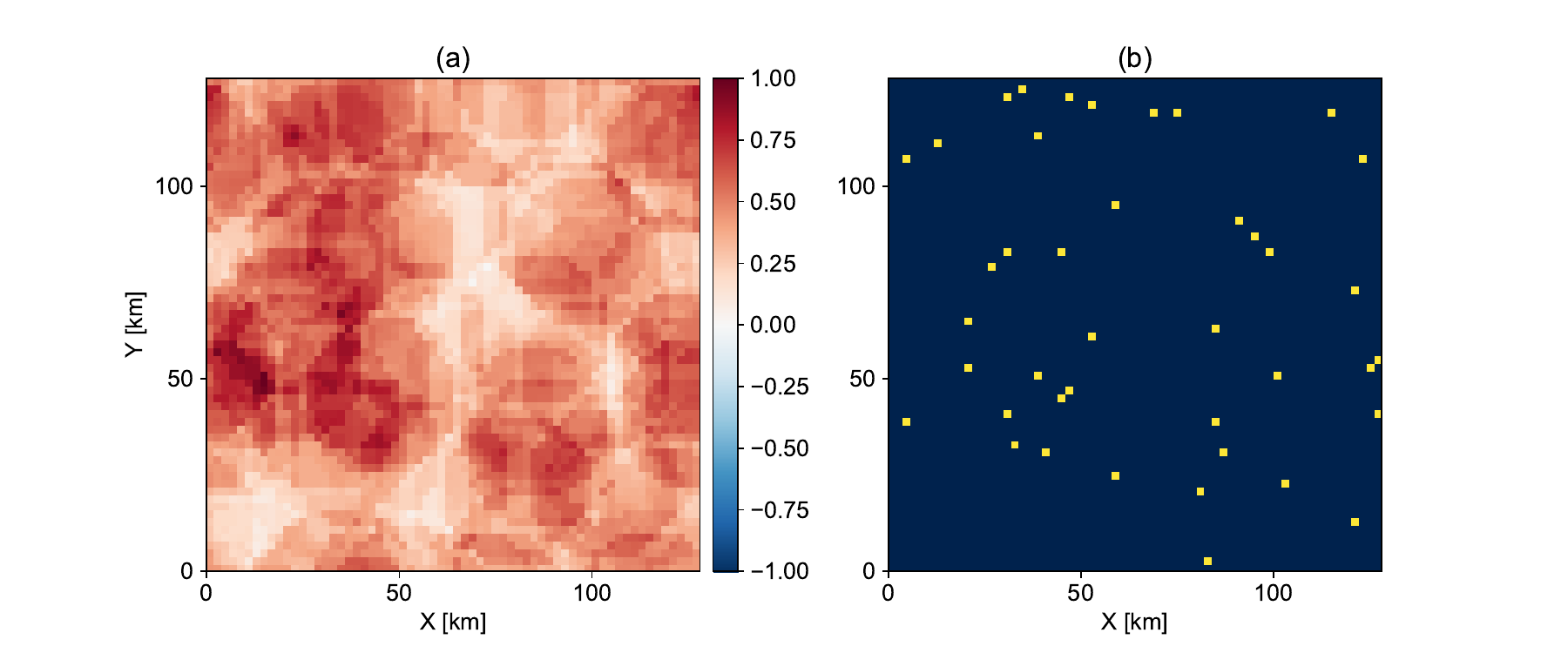}
    \caption{Snapshots of the horizontal distribution of (a) $J$ and (b) $I$ at t = 300 with a smaller domain $L=\SI{128}{km}$.}
    \label{imshow-small}
  \end{center}
\end{figure}

While keeping $f_s$ at -0.0045, which is the subsidence forcing that produced an aggregated convection state with $L=\SI{256}{km}$, the domain size is reduced to $\SI{128}{km}$, which is the half of the original. The other parameters are identical to Table \ref{tab_param}. With this configuration, the quasi-steady state of $J$ changes to a scattered case (Fig. \ref{imshow-small}). This behavior is consistent with the results observed in the RCE experiments that the self-aggregation is less likely to occur when the region size is smaller \citep{Yanase_2020, Hung_2021, Muller_2012}.

\begin{figure}
  \begin{center}
    \includegraphics[width=120mm]{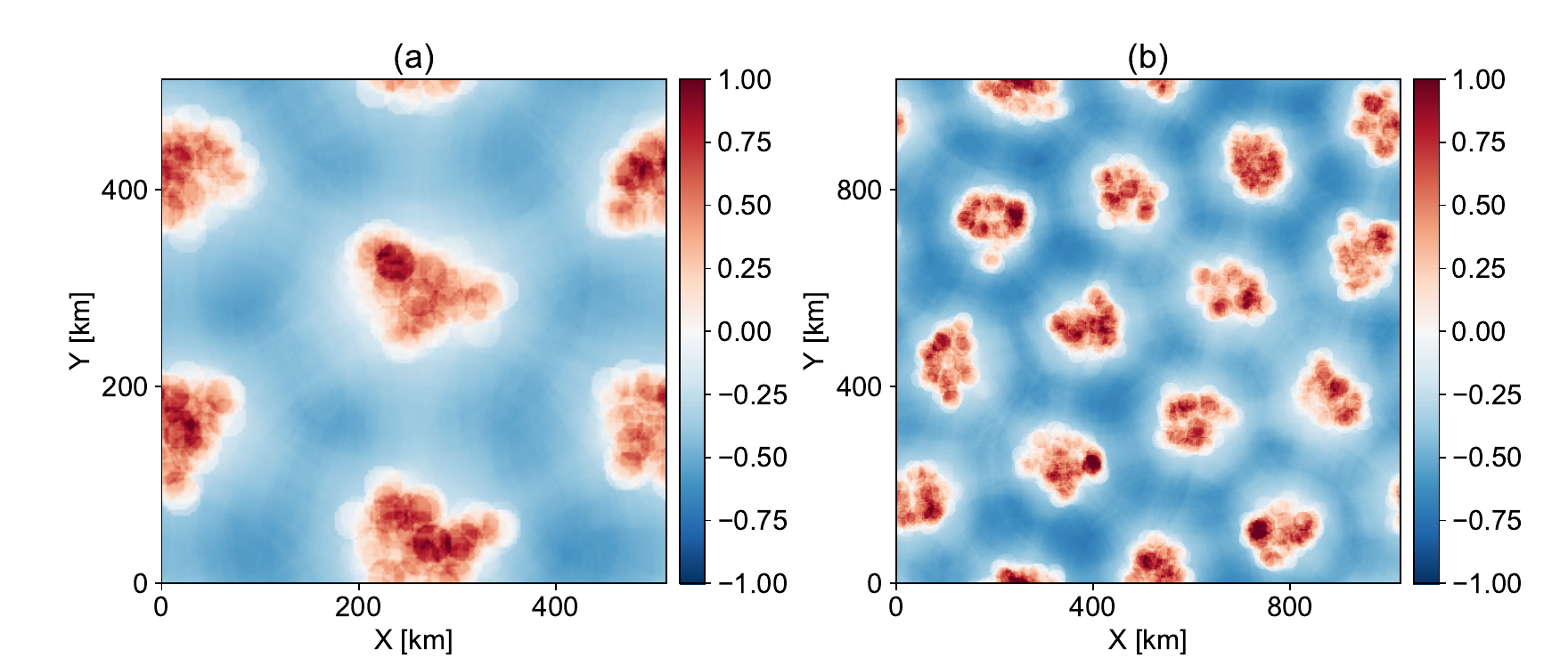}
    \caption{Snapshots of the horizontal distribution of $J$ at $t=300$ with larger domain sizes, (a) $L=\ $\SI{512}{km} and (b) $L=\ $\SI{1024}{km}. In these cases, the strength of interaction between clouds is truncated at \SI{200}{km}, and assumed to be zero beyond this distance.}
    \label{imshow-large}
  \end{center}
\end{figure}

Conversely, a set of experiments in larger model domains with sizes $L=\SI{512}{km}$ and $\SI{1024}{km}$ are conducted. For these cases, we limit the reach of the effect of subsidence to a finite distance following the interaction profile adopted in \citet{Randall_1980}. Here, the application $F(\mathbf{r},\mathbf{r}_i) = -0.0045$ is limited from $\SI{20}{km}$ to $\SI{200}{km}$. $f_s$ outside of the $\SI{200}{km}$ circle is set to zero, representing no interaction. The snapshots at $t=300$ hours are displayed in Fig. \ref{imshow-large}. 
There are multiple clusters of clouds that do not interfere with each other, and they are surrounded by subsidence regions in honeycomb pattern. The pattern formation is presumably related to setting a characteristic length scale of the effect of subsidence.
This result is similar to that of \citet{Randall_1980}, who suggested that the pattern formation of clouds and moisture is strongly influenced by the horizontal profile of the cloud interaction. 
The spatial distribution of the cloud clusters resembles that of open cells of shallow convection presented by \citet{Feingold_2010} using LES, or that of multiple cloud clusters in the RCE simulation with extremely large domains \citep{Wing_2020, Yanase_2022}. Interestingly, such hexagonal or mesh-like structures are not limited to atmospheric phenomena. They are universally observed in a variety of self-organizing systems such as Bénard convection \citep{Benard_1901}, buckling of metals \citep{Carlson_1967}, and chemical reaction of chlorite-iodide-malonic acid \citep{Ouyang_1991}.

\begin{figure}
  \begin{center}
    \includegraphics[width=120mm]{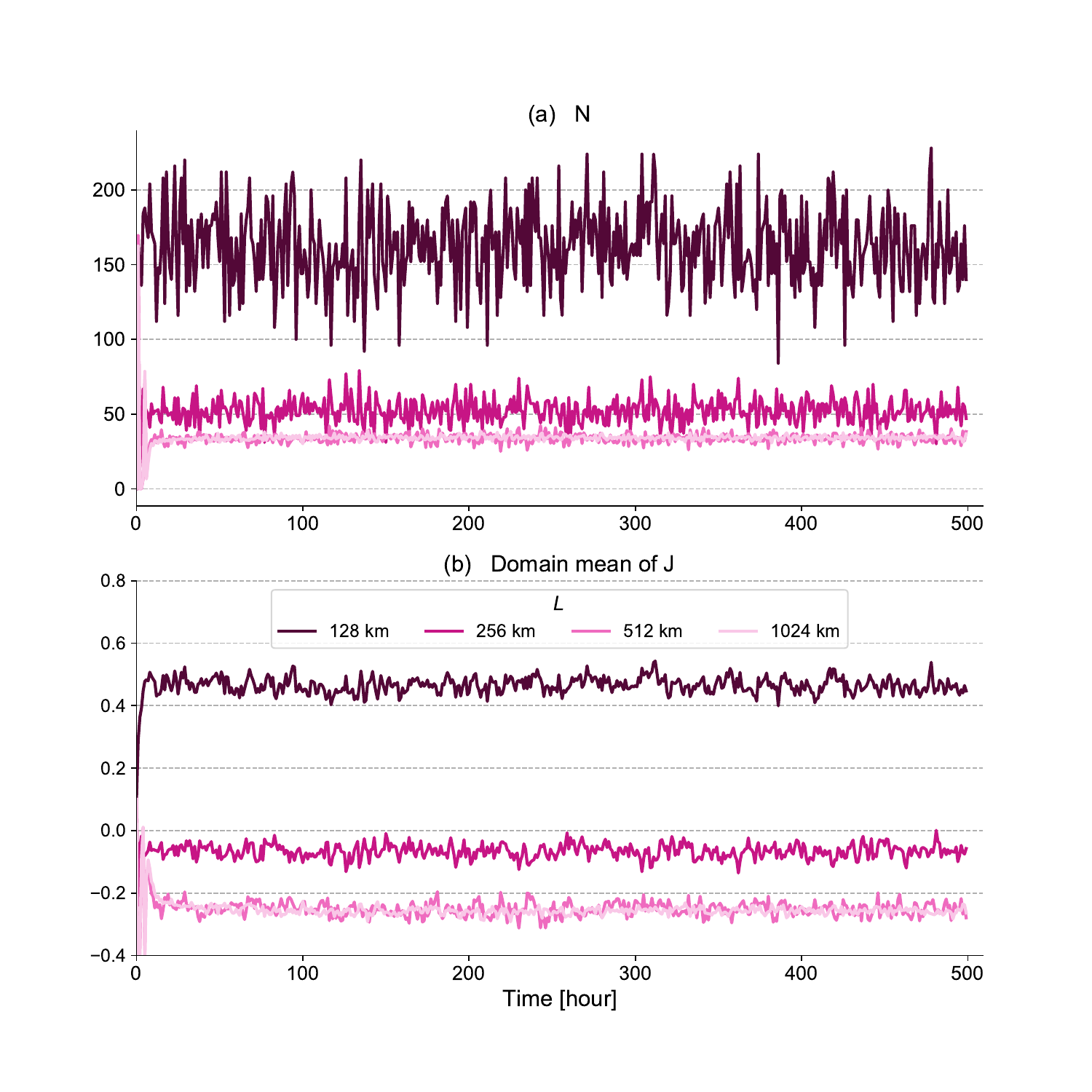}
    \caption{Same as Fig. \ref{JN-time}, but for the simulations with different domain sizes. $N$ is rescaled to match the relative area of the experimental domain compared to the $L=\ $\SI{256}{km} simulation.}
    \label{JN-time-L}
  \end{center}
\end{figure}

Figure \ref{JN-time-L} shows the time evolution of $N$ and the domain mean of $J$ for each domain size. As in the case with varying $f_s$, there is a clear difference between the regimes where convection is aggregated or not. For the case of scattered convection ($L=\SI{128}{km}$), the number of convective grid points lies in the range of approximately 150 to 170, similar to the cases with $f_s=-0.0005$ and $-0.0015$ in the $L=\ $\SI{256}{km} experiments, and the domain-mean $J$ remains positive. In the aggregated cases ($L=\ $\SI{256}{km}, \SI{512}{km}, and \SI{1024}{km}) the number of convective grid points is less than half of that in the $L=\SI{128}{km}$ case, and the domain-mean $J$ remains negative.

\subsection{Stability analysis of the steady states}
\label{stability}
\begin{figure}
  \begin{center}
    \includegraphics[width=0.65\linewidth]{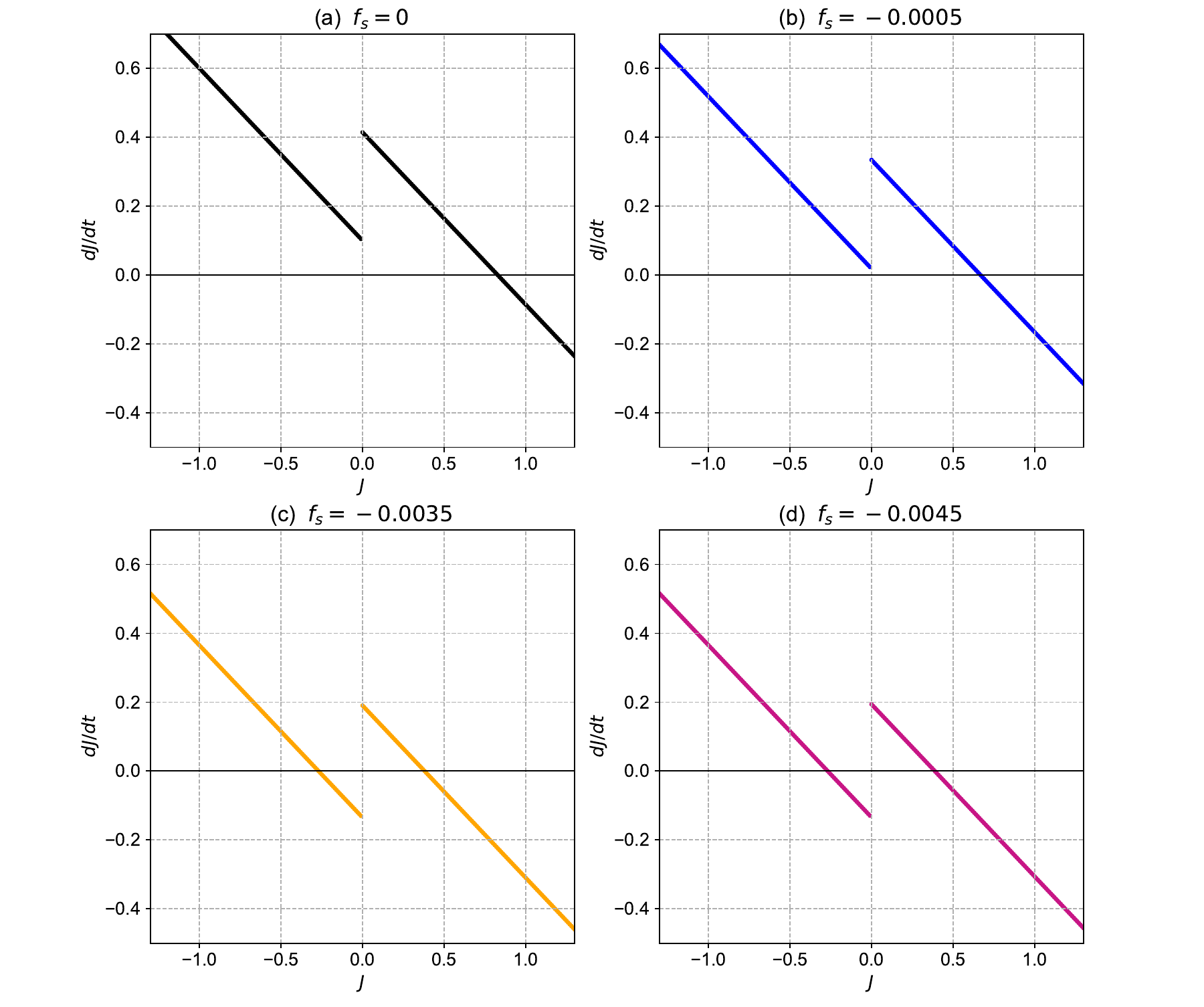}
    \caption{$\frac{dJ}{dt}$ as a function of $J$ for the four cases (a) $f_s=-0$, (b) $f_s=-0.0005$, (c) $f_s=-0.0035$, and (d) $f_s=-0.0045$.}
    \label{Jdot}
  \end{center}

  \begin{center}
    \includegraphics[width=0.65\linewidth]{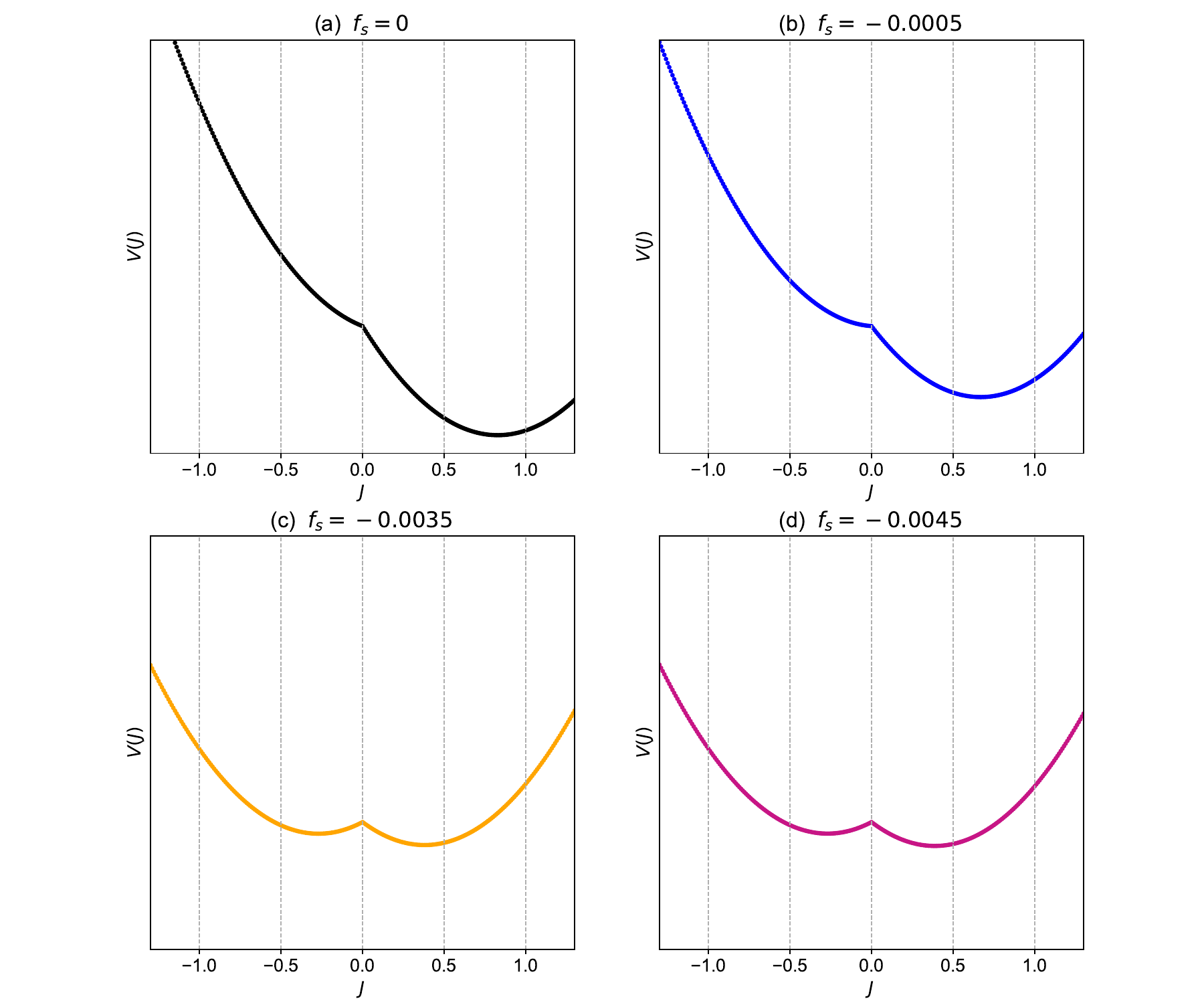}
    \caption{The potential $V$ as a function of $J$ for the four cases (a) $f_s=-0$, (b) $f_s=-0.0005$, (c) $f_s=-0.0035$, and (d) $f_s=-0.0045$.}
    \label{potential}
  \end{center}
\end{figure}

In this subsection, a stability analysis for the steady solutions of $J$ is conducted. We examine if the persistent separation of the moist and dry regions are stable in terms of dynamical systems. Taking the horizontal profile of the interaction between clouds (Eq. (\ref{eq_f})) into account, the prognostic equation for $J$ may be rewritten as
\begin{equation}
  \begin{split}
    \label{eq_J1}
    \frac{dJ(\mathbf{r})}{dt}
    &= f_c n_c(\mathbf{r}) + f_s(N-n_c(\mathbf{r})) + \alpha - \frac{1}{\tau_d} J(\mathbf{r}) \\
    &= - \frac{1}{\tau_d} J(\mathbf{r}) + (f_c - f_s)n_c(\mathbf{r}) + \alpha + f_sN, \\
  \end{split}
\end{equation}
where $n_c(\mathbf{r})$ denotes the number of convective grid points within a circle of the 20-km radius centered at the position $\mathbf{r}$. Since Eq. (\ref{eq_J1}) involves the local variable $n_c(\mathbf{r})$ and the large-scale variable $N$ on the dynamics of $J(\mathbf{r})$, a few assumptions are additionally made to simplify the problem and to classify the number of stable solutions for $J(\mathbf{r})$. Here, we neglect the local variability of $J$ in the vicinity of the position $\mathbf{r}$ and assume that $J$ takes a uniform value $J(\mathbf{r})$ within the circle of \SI{20}{km} radius centered at $\mathbf{r}$. By this assumption, the expected value of $n_c(\mathbf{r})$ can be described in terms of $J(\mathbf{r})$ as
\begin{equation}
    \label{nc}
  n_c(\mathbf{r}) \simeq \frac{ap_{\gamma}}{dx^2} Y(J(\mathbf{r})),
\end{equation}
where $a$ is the area of a cold pool with the maximum radius, and $Y$ is a Heaviside step function. Thus, an approximation of Eq. (\ref{eq_J1}) may be written as
\begin{equation}
  \label{eq_J2}
    \frac{dJ(\mathbf{r})}{dt} \simeq - \frac{1}{\tau_d} J(\mathbf{r}) + \frac{(f_c-f_s)ap_{\gamma}}{dx^2} Y(J(\mathbf{r})) + \alpha + f_s N.
\end{equation}
Using this form, the tendency of $J$ (left-hand side) is evaluated as a function of $J$, and it is displayed in Fig. \ref{Jdot} for four different values of $f_s$. The parameters except for $f_s$ are the same as Table \ref{tab_param}. $N$ is estimated by the time average in each simulation in Section \ref{result}\ref{var_fs}. The case of $f_s=0$ is added as another example of scattered convection. The steady solutions of Eq. (\ref{eq_J2}) that satisfy the condition $\frac{\partial}{\partial J}(\frac{dJ}{dt}) < 0$ are stable by definition. The scattered cases (Fig. \ref{Jdot}a, b) have a single stable solution with $J>0$, while the aggregated cases (Fig. \ref{Jdot}c, d) have two stable solutions for moist ($J>0$) and dry ($J<0$) states.

The form of corresponding potentials for the dynamics of $J$ are obtained by integrating Eq. (\ref{eq_J2}) with respect to $J$ (Fig. \ref{potential}). The potential of the current model is expressed by two quadratic functions connected at $J=0$. The scattered cases have a single equilibrium (Fig. \ref{potential}a, b). On the other hand, the aggregated cases have a double-well potential (Fig. \ref{potential}c, d), in which the perturbation of $J$ is amplified near the boundary of dry and moist regimes as illustrated in Fig. \ref{stab-schem}.

Whether the self-aggregation occurs with a given set of the parameters is estimated by taking the summation of Eq. (\ref{eq_J1}) over the whole domain, which equals to
\begin{equation}
  \label{eq_sum}
  \displaystyle\sum_{\mathbf{r}} \frac{dJ(\mathbf{r})}{dt} = - \frac{A}{dx^2}\frac{\overline{J}}{\tau_d} + f_c\frac{a}{dx^2}N + f_s\frac{A-a}{dx^2}N + \frac{A}{dx^2}\alpha,
\end{equation}
where $A$ is the area of the model domain, and $\overline{J}$ is the domain mean of $J$.
For the case with the scattered convection, $N=\frac{p_{\gamma}A}{dx^2}$ and $\overline{J}>0$ are reasonably expected. Then, the necessary condition for the scattered regime is
\begin{equation}
  \label{eq_Jcond}
    - \frac{\overline{J}}{\tau_d} + \left[f_c\frac{a}{A} + f_s\left(1-\frac{a}{A}\right)\right]\frac{p_{\gamma}A}{dx^2} + \alpha = 0 \ \ (\overline{J}>0).
\end{equation}
When $f_s$ is free and the other parameters are fixed as in Table \ref{tab_param}, the condition of $f_s$ for the scattered convection can be computed as
\begin{equation}
  \label{eq_fscond}
    f_s > -\frac{1}{1-\frac{a}{A}}\left( f_c \frac{a}{A} + \frac{\alpha dx^2}{p_{\gamma}A} \right) = -0.0026.
\end{equation}
Although this critical value of $f_s$ is somewhat smaller than that of the simulations results (Fig. \ref{J-pdf}b), the above estimation successfully explains the results that smaller values of $f_s$ is favorable for the aggregated state.

Similarly, the critical values can be derived from Eq. (\ref{eq_J2}) when $f_s$ is fixed and one of the other parameters are free. For instance, the larger values of the size of the cold pools $a$ introduced in Eq. (\ref{nc}) is favorable for the scattered state, while the larger values of the domain size $A$ in Eq. (\ref{eq_sum}) is favorable for the aggregated state. These results are consistent with the current lattice model simulation in Section \ref{result}\ref{var_a} and Section \ref{result}\ref{var_A}, respectively.

\section{Conclusions and discussions}
\label{conclusion}
In this paper, a simple stochastic two-dimensional lattice model for the pattern formation of a deep convection ensemble has been developed. 
A series of simulations was conducted by changing the magnitude of the effect of subsidence enhanced around convection. The results showed that the strong contrast between the promotion of convection in the vicinity of the existing convection and the suppression elsewhere caused a heterogeneous spatial distribution of convective cells in the quasi-steady state, which resembled the aggregated cloud clusters in the RCE simulations and the cloud fields in the real atmosphere. The model also successfully explained a variety of spatial patterns of cloud field according to the relationship between the geometrical structure of the model domain and the length scale of the effect of subsidence to reach. 

One of the limitations of this lattice model is that the number of convective cells over the entire domain in quasi-equilibrium state cannot be determined from large scale parameters. This makes it difficult to treat the cumulus ensemble as a canonical ensemble described in equilibrium statistical mechanics. In addition, since this model has a time scale of dissipation of two hours, the oscillatory behavior on the time scale of tens of days observed in idealized experiments using CRMs \citep{Patrizio_2019} are not incorporated. The multi-scale nature of the dynamics of cloud systems should be pursued in the future studies.

%

%

\clearpage
\acknowledgments
This work was supported by JSPS KAKENHI Grant Number 23K25939, 20H05731, 20J11246. The authors have no conflicts to disclose.

%
%
\datastatement
All code used in this study will be made available on GitHub after completing the peer review process for this article.

%






%



\bibliographystyle{ametsocV6}
\bibliography{0-main}

\end{document}